\documentclass[aps,prd,reprint,amsmath,amssymb,preprintnumbers,superscriptaddress,nofootinbib]{revtex4-1}
\usepackage[colorlinks]{hyperref}
\usepackage{graphicx}
\usepackage{comment}
\newcommand{\tx}{\text}
\newcommand{\ov}{\over}
\newcommand{\p}{\partial}
\newcommand{\J}{\tx{J}} % Jordan
\newcommand{\MP}{M_\text{P}}

\newcommand{\paren}[1]{\left(#1\right)}
\newcommand{\fn}[1]{\!\left(#1\right)}

\newcommand{\ab}[1]{\left|#1\right|}
\newcommand{\df}{\text{d}}

\newcommand{\mc}{\mathcal}

\newcommand{\GeV}{\,\text{GeV}}

\begin{document}
%%%%%%%%%%%%%%%%%%%%%%%%%%%%%%%%%%%%%%%%%%%%%%%%%%%%%%%

\preprint{CTPU-17-18}
\preprint{OU-HET/935}
\title{
\Large
Hillclimbing Higgs inflation
}
\renewcommand{\thefootnote}{\alph{footnote}}

\author{
Ryusuke Jinno
}
\author{
Kunio Kaneta
}

\affiliation{
Center for Theoretical Physics of the Universe, Institute for Basic Science (IBS), Daejeon 34051, Korea
}

\author{
Kin-ya Oda
}

\affiliation{
Department of Physics, Osaka University, Osaka 560-0043, Japan\\
}

\begin{abstract}
We propose a realization of cosmic inflation with the Higgs field when the Higgs potential has degenerate vacua by employing the recently proposed idea of hillclimbing inflation. The resultant inflationary predictions exhibit a sizable deviation from those of the ordinary Higgs inflation.
\end{abstract}

\maketitle

%%%%%%%%%%%%%%%%%%%%%%%%%%%%%%%%%%%%%%%%%%%%%%%%%%%%%%%
\section{Introduction}
\label{sec:Introduction}
%%%%%%%%%%%%%%%%%%%%%%%%%%%%%%%%%%%%%%%%%%%%%%%%%%%%%%%

Inflation plays an essential role in modern cosmology~\cite{Starobinsky:1980te,Sato:1980yn,Guth:1980zm},
not only by addressing the horizon and flatness problems~\cite{Guth:1980zm},
but also by giving primordial seeds for late-time structures~\cite{Mukhanov:1981xt}.
The properties of primordial perturbations have been strongly constrained by precision cosmology, 
especially by the cosmic microwave background (CMB) observations~\cite{Ade:2015lrj},
and such observations are expected to explore the inflationary physics much further in the forthcoming decade.
Nevertheless, the identity of the inflaton, the scalar field causing inflation, is still veiled in mystery.

The Higgs particle---the quantum fluctuation of the Higgs field around its potential minimum---had 
long been the last missing element of the Standard Model (SM), and was finally  
discovered in 2012~\cite{Aad:2012tfa,Chatrchyan:2012xdj}.
Ever since, the Higgs field has been the only (possibly) elementary scalar field observed by human beings.
The possibility of realizing inflation with this Higgs field has been studied extensively,
and it has turned out that the Higgs field can indeed be identified as the inflaton
with the help of a large non-minimal coupling $\xi \sim 10^5$ to the Ricci scalar~\cite{Bezrukov:2007ep}.\footnote{
In earlier Ref.~\cite{Salopek:1988qh}, the Higgs inflation with essentially the same parameters $\xi \sim 10^4$ 
and $\lambda \sim \paren{\xi/10^5}^2 \sim 10^{-2}$ has also been sketched; see also Refs.~\cite{Lucchin:1985ip,Futamase:1987ua,vanderBij:1993hx,vanderBij:1994bv,CervantesCota:1995tz}.
It is noted that we may also cope with a smaller $\xi\sim10$--$10^2$ under the SM criticality~\cite{Hamada:2014iga,Bezrukov:2014bra,Hamada:2014wna}.
See also Ref.~\cite{Ema:2016dny} for the explosive production of longitudinal gauge bosons 
and possible strong coupling issues under the presence of the large non-minimal coupling.
}
This scenario, now called the Higgs inflation, 
has been found to fit in the most favored region by CMB observations~\cite{Ade:2015lrj}.

Regarding the mass of the Higgs particle $m_H = 125.09 \pm 0.24\,\text{GeV}$~\cite{Aad:2015zhl},
there was an interesting prediction based on the Multiple Point Principle (MPP)~\cite{Froggatt:1995rt}.\footnote{See Appendix~D in Ref.~\cite{Hamada:2015ria} for a review, and Ref.~\cite{Nielsen:2012pu} for possible generalizations.}
The MPP requires that there exist another vacuum in the Higgs potential 
around the Planck scale, in addition to the electroweak one.
This means that the Higgs quartic coupling and its beta function both vanish there, 
$\lambda\sim\beta_\lambda\sim0$.\footnote{
See e.g.\ Refs.~\cite{Hamada:2012bp,Buttazzo:2013uya} for more recent analyses.
Especially, it is intriguing that the bare Higgs mass can also vanish around the Planck scale, 
and hence there can be a triple coincidence~\cite{Hamada:2012bp}.
}
The observed Higgs mass has turned out to be almost within 1$\sigma$ 
from the value $m_H=135 \pm 9\GeV$ predicted in this way~\cite{Froggatt:1995rt}.

Even though these two scenarios, the Higgs inflation and the MPP, seem attractive, 
difficulties arise when it comes to combining them.
The MPP requires a degenerate vacuum around the Planck scale, 
which spoils the monotonicity of the Higgs potential 
which is necessary for a successful inflation~\cite{Hamada:2013mya}. 
On this regard, an interesting proposal has recently been made by two of the present authors:
the hillclimbing inflation~\cite{Jinno:2017jxc}.
This is a general framework which enables a successful inflation with an inflaton potential 
with multiple vacua.
This idea opens up a new possibility of 
identifying the Higgs field as the inflaton
while having degenerate vacua in the Higgs potential.
The aim of this paper is to pursue this possibility.\footnote{
The gauge-Higgs unification models fit in the periodic potential case in the general consideration of the hillclimbing inflation~\cite{Jinno:2017jxc}. Such a possibility will be pursued in a separate publication.
}

This paper is organized as follows.
In Sec.~\ref{sec:General} we briefly summarize 
inflationary behavior and predictions in the hillclimbing inflation.
Then in Sec.~\ref{sec:HillclimbingHiggs} we propose an inflation model 
using the Higgs field as the inflaton.
We conclude in Sec.~\ref{sec:Conc}.

%%%%%%%%%%%%%%%%%%%%%%%%%%%%%%%%%%%%%%%%%%%%%%%%%%%%%%%
\section{Hillclimbing inflation and its predictions}
\label{sec:General}
%%%%%%%%%%%%%%%%%%%%%%%%%%%%%%%%%%%%%%%%%%%%%%%%%%%%%%%

In this section, we briefly summarize 
the inflaton behavior and inflationary predictions in the general hillclimbing inflation.
We start from the Jordan-frame action that has a non-minimal coupling between the inflaton and gravity:
\begin{align}
S
&=
\int \df^4x \sqrt{-g_\J} 
\left[
\frac{1}{2} \Omega R_\J
- \frac{1}{2} g_\J^{\mu \nu} \partial_\mu \phi_\J \partial_\nu \phi_\J
- V_\J
\right],
\label{eq:SJ}
\end{align}
where (and throughout the paper) we work in the Planck units $\MP = 1/\sqrt{8\pi G} = 1$ unless otherwise stated;
the subscript J indicates that the quantity is given in the Jordan frame; 
$\phi_\J$, $R_\J$ and $V_\J(\phi_\J)$ are 
the inflaton, the Ricci scalar and the inflaton potential, respectively;
and
we assume that the conformal factor $\Omega(\phi_\J)$ is positive for the inflaton field values we consider.
Under the Weyl rescaling $g_{\mu \nu} = \Omega g_{\J \mu \nu}$, the Ricci scalar transforms as
\begin{align}
R_\J
&=
\Omega 
\left[ 
R + 3 \Box \ln \Omega 
- \frac{3}{2} \paren{\partial \ln \Omega}^2
\right],
\label{eq:R}
\end{align}
and we obtain the Einstein-frame action that has a canonically normalized Ricci scalar:
\begin{align}
S
&=
\int \df^4x \sqrt{-g} 
\left[
\frac{1}{2}R
- \frac{K}{2} \paren{\partial \phi_\J}^2 - V
\right],
\label{eq:SE}
\end{align}
where
\begin{align}
K
&= 
\frac{1}{\Omega}
+
\frac{3}{2}
\left(
\frac{\df\ln\Omega}{\df\phi_\J}
\right)^2,
\label{eq:KE}
\end{align}
$\paren{\p\phi_\J}^2=g^{\mu\nu}\p_\mu\phi_\J\p_\nu\phi_\J$,
and the Einstein-frame potential reads $V = V_\J / \Omega^2$.
If the second term dominates in Eq.~(\ref{eq:KE}),
the kinetic term in the action (\ref{eq:SE}) reduces to~\cite{Kallosh:2013tua}
\begin{align}
&
-\frac{K}{2}(\partial \phi_\J)^2 \simeq - \frac{3}{4} (\partial \ln \Omega)^2.
\label{eq:kinE}
\end{align}
This means that $\phi \simeq \sqrt{3/2}\ln \Omega$ works as a canonically normalized inflaton.
If the potential can be expanded as a series of $\Omega$, 
it will become exponentially flat in terms of $\ln\Omega$ as we will see below.

In Ref.~\cite{Jinno:2017jxc} it has been proposed to use $\Omega \ll 1$ for the inflationary dynamics,
instead of $\Omega \gg 1$.
In this class of models, called the hillclimbing inflation, 
it is assumed that $V_\J$ and $\Omega$ vanish at the same point $\phi_{\J *}$ in the field space,
and the Taylor expansion of the potential starts from 
$\Omega^n$ $(n \geq 2)$ around $\phi_\J = \phi_{\J *}$:
\begin{align}
V_\J
&= \sum_{k=n}^\infty \mc V_{\J,k}\,\Omega^k,
\label{eq:VJExpansion}
\end{align}
with $\mc V_{\J,k}$ being constants.
One of the examples is the case where the potential has a quadratic (or higher) minimum 
with vanishing cosmological constant at $\phi_\J = \phi_{\J  *}$,
and $\Omega$ vanishes linearly in $\phi_\J - \phi_{\J *}$ there.\footnote{
In fact, we have $\df V_\J/\df \Omega = (\df V_\J/\df \phi_J)/(\df \Omega/\df \phi_J) = 0$ at the minimum in this case.
However, it should be noted that this is just one of the possible realizations
and that 
the expansion (\ref{eq:VJExpansion}) is possible 
even if $\df \Omega/\df \phi_\J = 0$ at the vanishing point $\phi_\J = \phi_{\J *}$.
Also, if we allow $\Omega$ to be proportional to a fractional power of $\phi_\J - \phi_{\J *}$ 
around the vanishing point,
it is even possible to remove the assumption that $V_\J$ at $\phi_\J = \phi_{\J *}$ is a local minimum.}
In Sec.~\ref{sec:HillclimbingHiggs} 
we seek for a realization of this inflationary setup with the Higgs field.\footnote{\label{new footnote}
In the context of MPP, the existence of another local minimum of $V_\J$ is an essential requirement,
as explained in Sec.~\ref{sec:Introduction}.
The vanishing $\Omega$ at the coincident point can be regarded as a realization of the generalized MPP discussed in Ref.~\cite{Nielsen:2012pu}.
}
With the Jordan-frame potential (\ref{eq:VJExpansion}), 
the corresponding Einstein-frame potential $V=V_\J/\Omega^2$ reads
\begin{align}
V
&= 
V_0 
\left( 1 - \sum_{k = n}^\infty \eta_k \Omega^k \right)
= 
V_0 
\left( 1 - \sum_{k = n}^\infty \eta_k e^{-k\ab{\ln \Omega}} \right)
\label{eq:Vexpansion}
\end{align}
at $\Omega<1$, where we have written $V_0:=\mc V_{\J,2}$ and $\eta_k:=-\mc V_{\J,k+2}/\mc V_{\J,2}$ and the leading exponent $n\geq1$ dominantly determines the inflationary predictions.\footnote{
Having $n\geq2$ means that we assume $\mc V_{\J,k}=0$ for $k=3,\dots,n+1$.
}
The last expression in Eq.~(\ref{eq:Vexpansion}) tells that the potential is exponentially flat 
for the canonical inflaton field.
In Sec.~\ref{sec:HillclimbingHiggs} we will see that the leading power depends on 
the explicit form of the conformal factor we take.

It is remarkable that the Einstein-frame potential $V=V_\J/\Omega^2$ has been lifted up by the small $\Omega$ and made monotonic, even around a local minimum of $V_\J$.
As is pointed out in Ref.~\cite{Jinno:2017jxc}, inflation at $\Omega \ll 1$ means that
the Jordan-frame potential $V_\J = \Omega^2 V$ increases in time,
that is, the inflaton climbs up the Jordan-frame potential hill.
This observation is crucial in making successful inflation with
inflaton potentials having multiple vacua, as stressed in that paper.
In Sec.~\ref{sec:HillclimbingHiggs} we propose taking the SM Higgs field as the inflaton.

For the inflationary predictions, this class of models show attractor behavior called $\eta$-attractor.\footnote{
It can be shown that these models share the inflationary predictions with some branch of 
$\alpha$-attractor~\cite{Kallosh:2013yoa,Kallosh:2014rga}
at the leading order in the $e$-folding~\cite{Jinno:2017jxc}.
However, there are several reasons to distinguish $\xi$- and $\eta$-attractors from $\alpha$-attractor:
First of all, their actions generically differ even after the Weyl transformation.
Second, the reheating and preheating processes depend on
the preferred frame in which the canonically normalized matter fields are introduced;
see Ref.~\cite{Ema:2016dny} for example.
Finally, such a distinction is important in constructing inflation models 
with particle-physics motivated potentials, as stressed in Ref.~\cite{Jinno:2017jxc}.
}
Following the standard procedure, the slow-roll parameters with the potential (\ref{eq:Vexpansion}) 
are obtained as
\begin{align}
\epsilon_V
&\equiv 
\frac{1}{2}
\left( \frac{V'}{V} \right)^2
\simeq 
\frac{3}{4}\frac{1}{n^2N^2},
\;\;\;\;
\eta_V
\equiv 
\frac{V''}{V}
\simeq 
- \frac{1}{N},
\label{eq:srp}
\end{align}
where we used the following expression for the $e$-folding number $N$:
\begin{align}
N
&\simeq 
\frac{3}{2}\frac{1}{n^2\eta_n} \frac{1}{\Omega^n}.
\label{eq:N}
\end{align}
The inflationary predictions at the leading order in $N$ become
\begin{align}
n_s
&\simeq 1 - \frac{2}{N},
\;\;\;\;
r
\simeq \frac{12}{n^2N^2},
\label{eq:nsr}
\end{align}
where $n_s$ and $r$ are the spectral index and the tensor-to-scalar ratio, respectively.

%%%%%%%%%%%%%%%%%%%%%%%%%%%%%%%%%%%%%%%%%%%%%%%%%%%%%%%
\section{Hillclimbing Higgs inflation}
\label{sec:HillclimbingHiggs}
%%%%%%%%%%%%%%%%%%%%%%%%%%%%%%%%%%%%%%%%%%%%%%%%%%%%%%%

Now let us take the Higgs field as the inflaton. We write its effective potential as
\begin{align}
V_\J(\phi_\J)
&=
\frac{1}{4}\lambda_\tx{eff}\fn{\phi_\J}\phi_\J^4.
\end{align}
Around $\phi_\J = M \sim 10^{17\tx{--}18}\GeV$, 
the effective coupling can be approximated by~\cite{Hamada:2014wna}
\begin{align}
\lambda_\tx{eff}(\phi_\J)
&= 
\lambda_\tx{min} 
+ \beta_2 \paren{\ln{\phi_\J \ov M}}^2
+ \beta_3 \paren{\ln{\phi_\J \ov M}}^3
+ \cdots,
\end{align}
where $\beta_2\simeq 2 \times 10^{-5}=:\beta_2^\tx{SM}$ in the SM~\cite{Hamada:2013mya}.
The cubic and higher order terms are loop-suppressed, $\beta_3, \dots \ll \beta_2$, and will be neglected hereafter. 

In the following we set $\lambda_\tx{min}=0$ so that the potential becomes zero at $\phi_\J = M$ by assuming the MPP. In the SM, this is realized with the top quark mass $m_t \simeq 171.4\GeV$ for the strong coupling $\alpha_s \simeq 0.1185$, leading to $M\simeq4\times10^{18}\GeV$~\cite{Hamada:2014wna}.\footnote{
The precisely-measured top mass $m_t^\tx{MC}=173.1\pm0.6$\,GeV~\cite{PDG2017}
is the Monte-Carlo mass, which is just a parameter in the Monte-Carlo code. 
The mass relevant to our discussion is the pole mass. 
A theorists' naive average gives a combined result $m_t^\tx{pole}=173.5\pm1.1$\,GeV~\cite{PDG2017},
and the MPP value of the top pole mass $171.4\GeV$ is within $1.9\,\sigma$ of this bound.
}
However, the precise values of the $\beta_2$ and $M$ that realize $\lambda_\tx{min}=0$ are altered by extra particles such as the heavy right-handed neutrinos and the Higgs-portal dark matter; see e.g.\ Refs.~\cite{Davoudiasl:2004be,Iso:2009ss,Haba:2013lga,Hamada:2014xka,Haba:2014sia,Hamada:2017sga}.
Therefore we take them as free parameters hereafter.

%%%%%%%%%%%%%%%%
\begin{figure}
\centerline{
\includegraphics[width=\columnwidth]{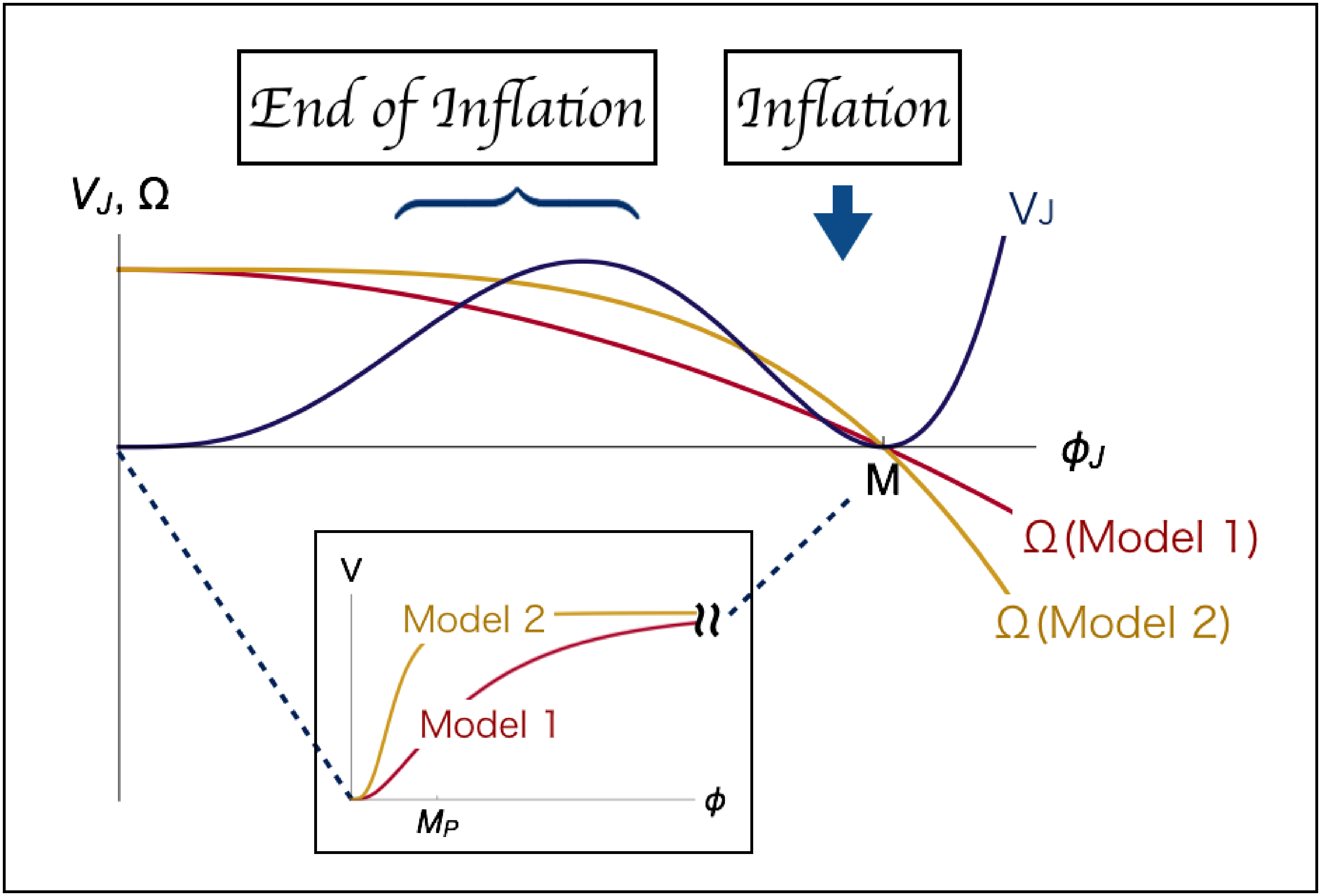}
}
\caption {\small
Illustration for the setup.
The Jordan-frame potential $V_\J$, shown in the blue line, has multiple vacua
at the electroweak scale $\sim v_{\rm EW}$ and the high scale denoted by $M \gg v_{\rm EW}$.
We assume that the conformal factor $\Omega$, denoted by the red or yellow lines for Model~1 and Model~2 in Eq.~\eqref{eq:OmegaModel}, respectively,
also vanishes at the point $\phi_\J = M$.
We also superimpose the Einstein-frame potential $V$ as a function of the canonically normalized field $\phi$.
The difference in the potential shape arises because
Model~1 corresponds to $n = 1$ while Model~2 corresponds $n = 2$ in Eq.~(\ref{eq:Vexpansion}).
In this figure the vertical axes is arbitrary, 
and we take $M = 0.1 \MP$.
}
\label{fig:pot}
\end{figure}
%%%%%%%%%%%%%%%%

%%%%%%%%%%%%%%%%
\begin{figure}
\centerline{
\includegraphics[width=\columnwidth]{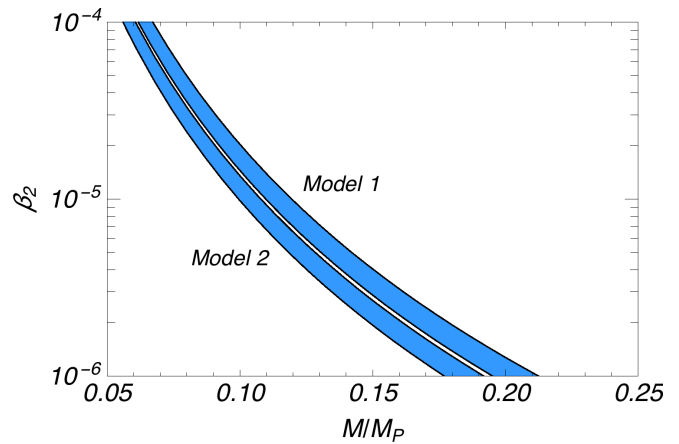}
}
\caption {\small
Parameter region which realizes the observed curvature perturbation
$A_s \simeq 2.2 \times 10^{-9}$.
The two bands correspond to Model~1 and Model~2 in Eq.~\eqref{eq:OmegaModel},
and the upper and lower lines for each band correspond to $N = 50$ and $60$, respectively.
See also Table~\ref{table}.
}
\label{fig:beta2}
\end{figure}
%%%%%%%%%%%%%%%%

Also, we consider the following forms for the conformal factor in this paper:\footnote{
We take $\Omega = 0$ at $\phi_\J = M$ as discussed in footnote~\ref{new footnote}.
Then Eq.~\eqref{eq:OmegaModel} is the two simplest possibilities that lead to $\Omega=1$ at $\phi_\J=0$ at the same time.
}
\begin{align}
\Omega
&=
\begin{cases}
\displaystyle 1 - \left( \frac{\phi_\J}{M} \right)^2
& {\rm (Model \; 1),} \\ 
\displaystyle 1 - \left( \frac{\phi_\J}{M} \right)^4
& {\rm (Model \; 2).} 
\end{cases}
\label{eq:OmegaModel}
\end{align}
We summarize the setup in Fig.~\ref{fig:pot}.
Given this setup, the Einstein-frame potential is expanded as
\begin{align}
V
&= 
\left\{
\begin{matrix}
\displaystyle \frac{\beta_2 M^4}{16} \left( 1 - \Omega - \cdots \right)
\;\;\;\;\;\;\;\;\;\;
&\;\;\;\; {\rm (Model \; 1)}, \\ 
\displaystyle \frac{\beta_2 M^4}{64} \left( 1 - \frac{1}{12}\Omega^2 - \cdots \right)
&\;\;\;\; {\rm (Model \; 2)}.
\end{matrix}
\right. 
\label{eq:VE}
\end{align}
Therefore, the leading exponent is given by $n=1$ and $2$ for Model~1 and 2, respectively,
and the potential height in the Einstein frame is given by $V_0 \sim \beta_2 M^4$.
Taking Eq.~(\ref{eq:srp}) and the curvature perturbation $A_s \sim V_0 / \epsilon_V$ into account,
one sees that the observed value $A_s \simeq 2.2 \times 10^{-9}$
constrains the model parameters along $M \propto \beta_2^{-1/4}$.
Figure~\ref{fig:beta2} shows such a constraint for each of Model~1 and 2.
The two bands correspond to Model~1 and 2,
and the upper and lower lines for each band correspond to $N = 50$ and $60$, respectively.
In making this figure we numerically solved for the $e$-folding $N$ under the slow-roll assumption, defining the end of inflation by ${\rm max}(\epsilon_V,\eta_V) = 1$.
It should be mentioned that while we have investigated only two simple models, there are various possible choices of $\Omega$ which gives different viable parameter spaces.
In addition, as mentioned above, the values of $\beta_2$ and $M$ may easily change in models beyond the SM by the existence of additional particles and associated intermediate scales; see e.g.\ Refs.~\cite{Davoudiasl:2004be,Iso:2009ss,Haba:2013lga,Hamada:2014xka,Haba:2014sia}.

%%%%%%%%%%%%%%%%
\begin{figure}
\centerline{
\includegraphics[width=0.9\columnwidth]{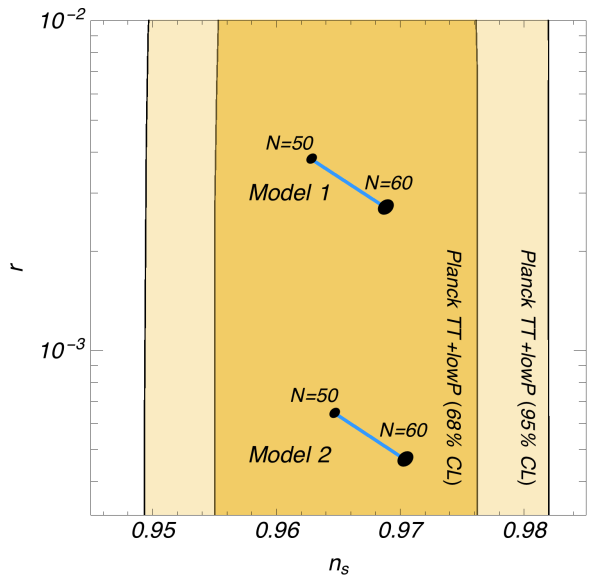}
}
\caption {\small
Inflationary predictions in the hillclimbing Higgs inflation.
The two lines correspond to Model~1 and Model~2 in Eq.~\eqref{eq:OmegaModel},
and the left and right endpoints correspond to $N = 50$ and $60$, respectively.
}
\label{fig:nsr}
\end{figure}
%%%%%%%%%%%%%%%%

Figure~\ref{fig:nsr} shows the inflationary predictions in the hillclimbing Higgs inflation.
It is seen that the prediction of the tensor-to-scalar ratio differs between Model~1 and 2
because of the difference in the leading exponent.
See also Table~\ref{table}.
These predictions fall within the planned sensitivities by near-future experiments such as the POLARBEAR-2~\cite{Inoue:2016jbg}, LiteBIRD~\cite{Matsumura:2013aja} and CORE~\cite{Delabrouille:2017rct},
and in Model 1 there is a good possibility of distinguishing from the ordinary Higgs inflation by these experiments.

Note that the prediction for $r$ differs from the rough estimate (\ref{eq:nsr}) by ${\mathcal O}(10)\%$.
This is because Eq.~(\ref{eq:nsr}) is derived by taking only the leading term in Eq.~(\ref{eq:VE}) into account,
while higher order terms can contribute to the inflaton dynamics
as the conformal factor grows towards the end of inflation.
Such a contribution is larger if the coefficient of the leading term is smaller,
and this is why Model~2 shows a larger deviation from Eq.~(\ref{eq:nsr}) compared to Model~1.

%%%%%%%%%%%%%%%%
\begin{table}[h]
\begin{tabular}{|c|c|c|}
\hline
$\Omega$ &  Model~1 & Model~2 \\
\hline
\hline
$M/\MP$ & $\;[0.1005,0.0923]\;$ & $\;[0.0907,0.0837]\;$ \\
\hline 
$\phi_{\J,{\rm end}}/\MP$ & $\;[0.0635,0.0583]\;$ & $\;[0.0562,0.0519]\;$ \\
\hline 
$\phi_{\J,{\rm CMB}}/\MP$ & $\;[0.0991,0.0912]\;$ & $\;[0.0854,0.0791]\;$ \\
\hline 
$n_s$ & $\;[0.9628,0.9688]\;$ & $\;[0.9647,0.9703]\;$ \\
\hline 
$r$ & $\;[0.00381,0.00272]\;$ & $\;[0.000646,0.000468]\;$ \\
\hline
\end{tabular}
\caption{\small
Allowed region and inflationary predictions for $\beta_2 = 2 \times 10^{-5}$.
The left and right values correspond to $N = 50$ and $60$, respectively.
Model~1 and Model~2 are given in Eq.~\eqref{eq:OmegaModel}.
Note that the allowed region of $M$ scales as $\beta_2^{-1/4}$ (see the main text).
Also, the values of $n_s$ and $r$ do not depend on $\beta_2$ significantly.\smallskip\\
}
\label{table}
\end{table}
%%%%%%%%%%%%%%%%

In Table~\ref{table} we summarize the allowed value for $M$ and corresponding inflationary predictions
for $\beta_2 = 2 \times 10^{-5}$.
One sees that $M \sim 0.1\,\MP$ is favored for this value of $\beta_2$ 
and also that $\phi_\J$ at the CMB scale corresponds to $\sim 0.01\,M$ away from the potential minimum at $\phi_\J=M$.

\vskip 0.5in

%%%%%%%%%%%%%%%%%%%%%%%%%%%%%%%%%%%%%%%%%%%%%%%%%%%%%%%
\section{Conclusion}
\label{sec:Conc}
%%%%%%%%%%%%%%%%%%%%%%%%%%%%%%%%%%%%%%%%%%%%%%%%%%%%%%%

In this paper we have proposed a realization of cosmic inflation 
using the Higgs field with degenerate vacua.
This realization utilizes the recently proposed idea of hillclimbing inflation~\cite{Jinno:2017jxc},
which is a general framework to enable a successful inflation using an inflaton potential 
with multiple vacua.
It has been shown that a successful inflation occurs while the inflaton is climbing up the potential hill
from the high-scale vacuum around the Planck scale to the electroweak vacuum,
and that the resulting inflationary predictions come well within the region 
favored by the CMB observations, while showing a sizable deviation from those of the ordinary Higgs inflation.

Though in this paper we have considered only the case where the Higgs field has degenerate vacua,
the original proposal in Ref.~\cite{Jinno:2017jxc} can work also when the Higgs potential becomes 
negative at some scale. Such a study will be presented in a separate publication.

%%%%%%%%%%%%%%%%%%%%%%%%%%%%%%%%%%%%%%%%%%%%%%%%%%%%%%%
\section*{Acknowledgments}
%%%%%%%%%%%%%%%%%%%%%%%%%%%%%%%%%%%%%%%%%%%%%%%%%%%%%%%
\noindent
The authors are grateful to Y.~Hamada, H.~Kawai, Y.~Nakanishi, T.~Onogi, and S.~Rusak for useful discussions.
The work of R.J. and K.K.\ is supported by IBS under the project code, IBS-R018-D1.
The work of K.O.\ is supported in part by JSPS KAKENHI Grant Nos.~16J06151 and 23104009, 15K05053.

\vspace{6cm}

%%%%%%%%%%%%%%%%%%%%%%%%%%%%%%%%%%%%%%%%%%%%%%%%%%%%%%%
%\bibliographystyle{TitleAndArxiv}
\bibliography{Bibliography}
%%%%%%%%%%%%%%%%%%%%%%%%%%%%%%%%%%%%%%%%%%%%%%%%%%%%%%%

%%%%%%%%%%%%%%%%%%%%%%%%%%%%%%%%%%%%%%%%%%%%%%%%%%%%%%%
\end{document}